\documentclass[sigconf]{acmart}

\usepackage{array}
\newcolumntype{L}[1]{>{\raggedright\arraybackslash}p{#1}}

\AtBeginDocument{%
  }

\copyrightyear{2026}
\acmYear{2026}
\setcopyright{cc}
\setcctype{by}
\acmConference[CHI EA '26]{Extended Abstracts of the 2026 CHI Conference on Human Factors in Computing Systems}{April 13--17, 2026}{Barcelona, Spain}
\acmBooktitle{Extended Abstracts of the 2026 CHI Conference on Human Factors in Computing Systems (CHI EA '26), April 13--17, 2026, Barcelona, Spain}
\acmDOI{10.1145/3772363.3799276}
\acmISBN{979-8-4007-2281-3/2026/04}

\begin{document}


\title{Probing More-Than-Human Representation in Crisis Resilience Planning: An HCI Researcher Perspective}

\author{Tram Thi Minh Tran}
\authornote{Both authors contributed equally to this research.}
\orcid{0000-0002-4958-2465}
\email{tram.tran@sydney.edu.au}
\affiliation{Design Lab, School of Architecture, Design and Planning,
  \institution{The University of Sydney}
  \city{Sydney}
  \state{NSW}
  \country{Australia}
}

\author{Adrian Wong}
\authornotemark[1]
\orcid{009-0008-7340-7107}
\email{adrian.w@sydney.edu.au}
\affiliation{Design Lab, School of Architecture, Design and Planning,
  \institution{The University of Sydney}
  \city{Sydney}
  \state{NSW}
  \country{Australia}
}

\author{Callum Parker}
\orcid{0000-0002-2173-9213}
\email{callum.parker@sydney.edu.au}
\affiliation{Design Lab, School of Architecture, Design and Planning,
  \institution{The University of Sydney}
  \city{Sydney}
  \state{NSW}
  \country{Australia}
}

\author{Carlos Alfredo Tirado Cortes}
\orcid{0000-0003-0626-0914}
\email{carlos.tiradocortes@sydney.edu.au}
\affiliation{Design Lab, School of Architecture, Design and Planning,
  \institution{The University of Sydney}
  \city{Sydney}
  \state{NSW}
  \country{Australia}
}

\author{Marius Hoggenmueller}
\orcid{0000-0002-8893-5729}
\email{marius.hoggenmueller@sydney.edu.au}
\affiliation{Design Lab, School of Architecture, Design and Planning,
  \institution{The University of Sydney}
  \city{Sydney}
  \state{NSW}
  \country{Australia}
}

\author{Soojeong Yoo}
\orcid{0000-0003-3681-6784}
\email{soojeong.yoo@sydney.edu.au}
\affiliation{Design Lab, School of Architecture, Design and Planning,
  \institution{The University of Sydney}
  \city{Sydney}
  \state{NSW}
  \country{Australia}
}

\author{Nate Zettna}
\orcid{0000-0002-9915-6309}
\email{nate.zettna@sydney.edu.au}
\affiliation{Business School,
  \institution{The University of Sydney}
  \city{Sydney}
  \state{NSW}
  \country{Australia}
}

\author{Joel Fredericks}
\orcid{0000-0002-9847-5527}
\email{joel.fredericks@sydney.edu.au}
\affiliation{Design Lab, School of Architecture, Design and Planning,
  \institution{The University of Sydney}
  \city{Sydney}
  \state{NSW}
  \country{Australia}
}

\renewcommand{\shortauthors}{Tran and Wong et al.}

\begin{abstract}
Crisis resilience planning raises urgent questions about how to include non-human species and ecological systems in participatory processes, which remain largely human-centred. This paper reports on a workshop with HCI researchers examining how more-than-human representation is approached in crisis contexts. The workshop combined scenario-based discussion with two design probes—a voice-based conversational agent and an immersive embodied prototype—to support sustained discussion of how emerging technologies shape engagement with non-human perspectives. Participants focused not on system usability, but on deliberating representational choices, such as voice, embodiment, and realism, and their potential role within participatory planning processes. The findings suggest that giving ‘voice’ to non-humans is not a neutral act of translation, but a design challenge that introduces tensions between legitimacy, authority, and authenticity. This paper provides empirical insight into how HCI researchers conceptualise more-than-human representation and positions crisis resilience planning as a critical site for examining AI- and immersion-mediated representation.
\end{abstract}




\maketitle


\section{Introduction}

Recent work in Human-Computer Interaction (HCI) has argued for extending interaction design beyond human users to account for animals, ecosystems, and other non-human agencies~\cite{coulton2019more, giaccardi2020technology, poikolainen2022towards}. This body of work has explored multispecies ethics, alternative ontologies, and responsibilities toward non-human actors as ways of challenging anthropocentric assumptions in computing~\cite{frauenberger2019entanglement,roudavski2025dingo,mancini2023responsible}. However, questions remain about how such perspectives are practically approached within concrete decision-making contexts.

Crisis resilience planning brings these questions into sharp focus~\cite{smith2025contemporary}. Non-human species and ecological systems in the natural environment are directly affected by events such as bushfires, floods, and habitat loss~\cite{hirsch2025climate}, yet participatory planning processes remain largely human-centred, despite growing work that highlights the importance of incorporating non-human perspectives~\cite{usanga2024giving, light2024more, tomitsch2025ai}. As interest in more-than-human design grows, an open challenge is not only how technologies might give voice to non-human perspectives~\cite{nicenboim2023designing}, but how HCI researchers conceptualise and justify such acts of representation. While HCI researchers are not typically decision-makers within planning processes, they play a key role in designing the tools and representations through which non-human perspectives are mediated within participatory planning contexts. These ways of reasoning matter, as they shape whether and which non-human perspectives are considered legitimate in high-stakes decision-making.

This paper reports on a workshop conducted as part of the OzCHI 2025 conference~\cite{adrian2025ozchiws}, which brought together fifteen HCI researchers with expertise spanning interaction design, immersive technologies, AI, and more-than-human research. The workshop pursued two objectives: first, to examine how non-human perspectives might be considered in participatory planning for crisis resilience; and second, to explore how emerging technologies may be used to articulate non-human perspectives, including their potential roles and limitations. To support these objectives, the workshop combined scenario-based discussion with two design probes—a voice-based conversational agent and an immersive embodied agent—both representing a koala in a bushfire scenario. While the probes render a koala perspective in a literal sense, the focus of this paper is on how that act of giving voice is understood and debated by HCI researchers.

The findings show that while meaningful articulation is possible without voice or embodiment, the introduction of AI- and immersive-mediated representation makes representational choices consequential, shaping perceptions of legitimacy, authority, and authenticity. The challenge lies not only in creating representations of non-humans, but in understanding how such representations shape conversations and decision-making in planning contexts. This work invites discussion and critique around more-than-human representation in crisis contexts.

\section{Methods}


Fifteen participants, ranging from PhD students to senior academics and design educators, were organised into three small groups to capture a diversity of expertise and seniority within the HCI community. Participants were selected for disciplinary expertise and research experience (see \autoref{appendix:experts}). All participants were familiar with research-oriented discussion and critical reflection on design, technology, and ethics. Data collection was conducted with participant consent and in accordance with the University of Sydney–approved human research ethics protocol (HE001620).



\subsection{Scenario-Based Discussion}
Within a bushfire scenario, groups were asked to select a more-than-human actor and discuss how that actor’s perspective might be approached or articulated in crisis resilience planning. This activity focused on reasoning grounded in participants’ prior experience and expertise, without technological intervention.

\subsection{Design Probes}
The probes consisted of speculative prototypes introduced to elicit discussion and reflection~\cite{gaver1999design, wallace2013making}. They were not treated as finished systems for evaluation, but as shared reference points through which participants discussed how immersive and AI-supported tools might shape engagement with non-human perspectives. The koala was selected as a focal more-than-human actor due to its vulnerability to bushfire events and its relevance to the workshop scenario.

The two probes were designed to support discussion across different modes of representation. Prior work has examined voice and embodiment as relational dimensions shaping how AI agents are interpreted, including in science fiction contexts~\cite{wong2025voice}. In our workshop, the voice-based probe emphasised conversational interaction~\cite{nicenboim2023designing, Wong2025BeyondHuman}, while the immersive embodiment probe supported spatial and visual experience~\cite{Beaver2022JustinBeaver}. 

\begin{figure*}[ht]
    \centering
    \includegraphics[width=0.75\linewidth]{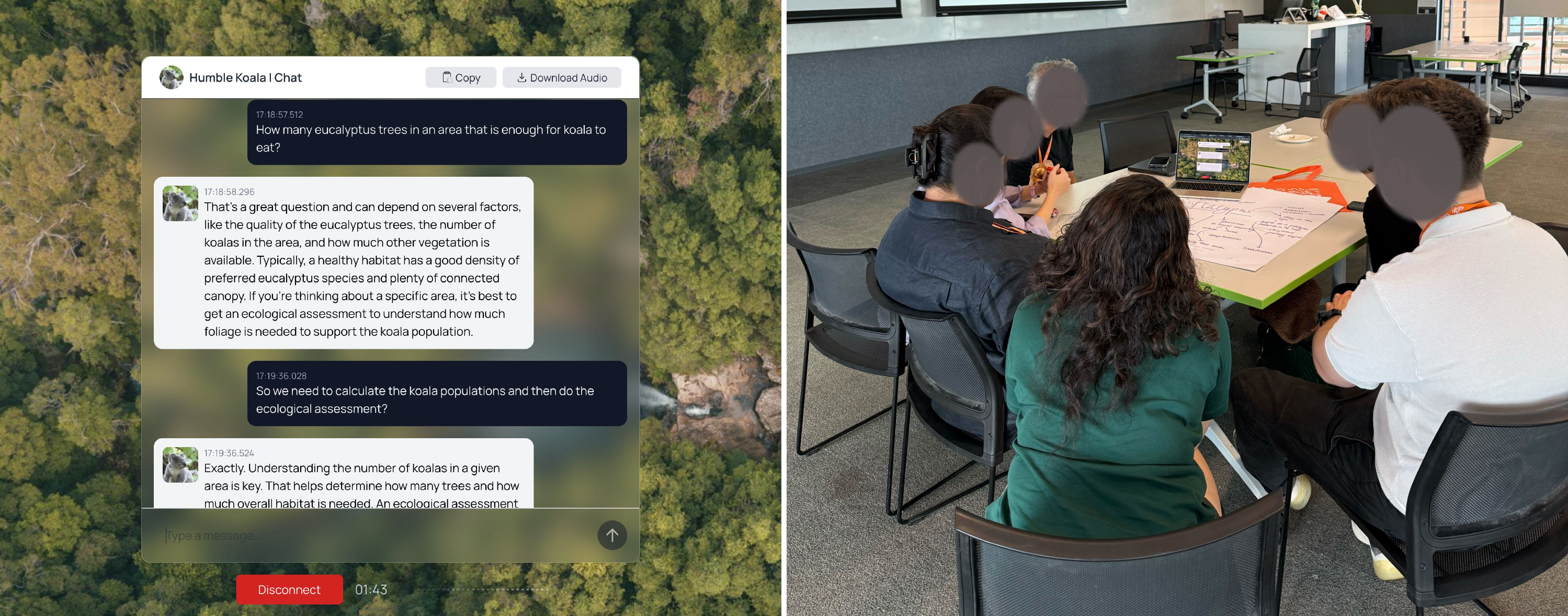}
    \caption{Voice-based conversational agent used as a design probe. (Left) The web-based interface of the koala conversational agent. (Right) Probe-supported group discussion, showing how the agent was explored collectively by participants during the workshop.}
    \Description{A voice-based conversational agent representing a koala, shown in a web interface. The screen displays a dialogue transcript while users speak questions and receive spoken responses. A group of participants interacts with the system around a shared laptop, taking turns speaking while others observe and discuss.}
    \label{fig:voice-agent}
\end{figure*}

\textit{Voice-based Conversational Agent}: 
Accessible via a website\footnote{\url{https://www.morethanhuman.me/agents}}, the agent represented a koala and was developed based on prior interviews with domain specialists. Its behaviour was authored through a structured system prompt defining identity, tone, behavioural constraints, and scope of knowledge, informed by thematic analysis of the expert interviews. Interaction was voice-based: participants could ask questions verbally, and responses were delivered through spoken output, with the conversation transcript displayed on the screen. Each group interacted with the probe for approximately 10–15 minutes, followed by a guided group discussion (see \autoref{fig:voice-agent}).

\begin{figure*}[ht]
    \centering
    \includegraphics[width=0.75\linewidth]{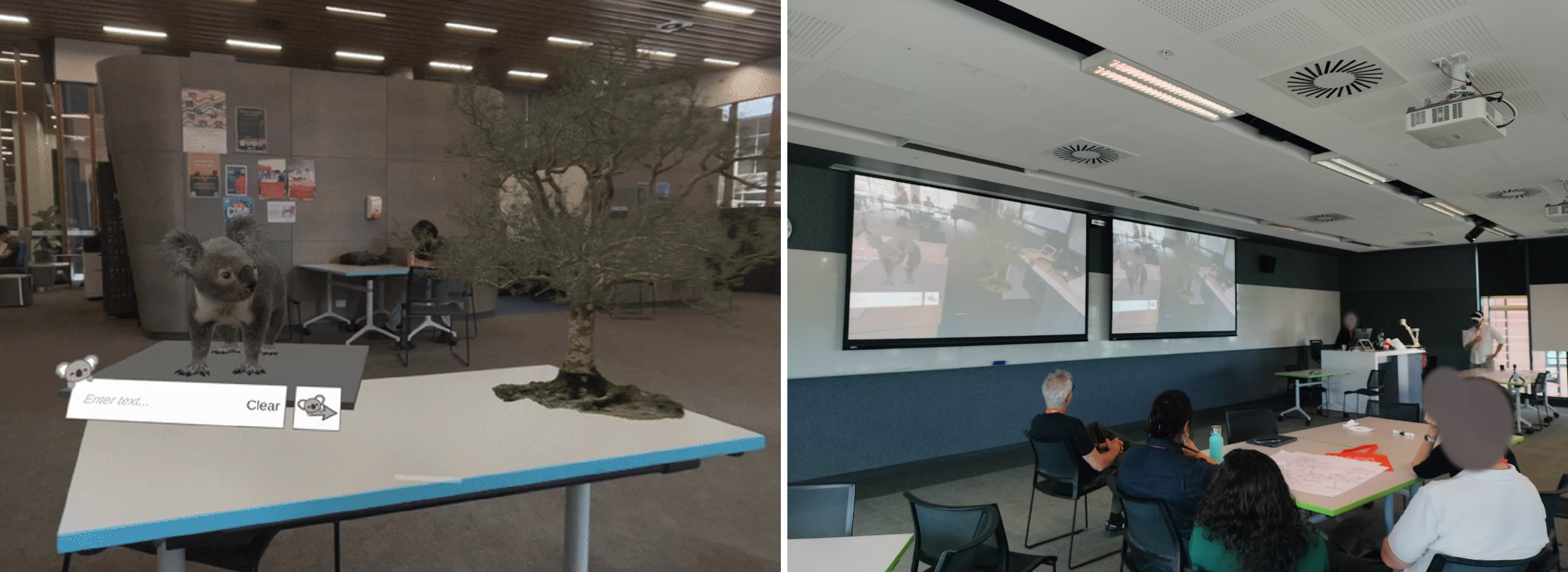}
    \caption{Immersive embodied koala agent used as a design probe. (Left) View from the Apple Vision Pro headset, showing the koala agent with visual and spatial embodiment in XR. (Right) Probe-supported group discussion, in which the headset view was projected to a shared screen while participants asked questions and reflected collectively.}
    \Description{An immersive embodied koala agent presented through an Apple Vision Pro headset. The headset view is projected onto a large screen, showing a 3D koala within a spatial scene alongside contextual visual information such as habitat elements. A researcher wears the headset while a group of participants watch, ask questions, and discuss the experience together.}
    \label{fig:xr-agent}
\end{figure*}

\textit{Immersive Embodied Prototype}: Presented using Apple Vision Pro, the same koala agent was rendered with visual and spatial embodiment. A researcher wore the headset, with their view projected onto a large screen for shared viewing. The agent responded to participant questions and presented contextual visual information in XR, such as preferred food trees. Participants collectively asked questions and reflected on the experience (see \autoref{fig:xr-agent}).


\subsection{Data Collection and Analysis}

Audio recordings were made for each group session and transcribed using OpenAI Whisper~\cite{openai_whisper}, with transcripts manually reviewed for accuracy. Groups also used flipcharts to externalise and organise their ideas throughout the workshop. Initial inductive coding was conducted by TTMT using collaborative FigJam boards to organise data into codes and iteratively develop themes. All authors then reviewed and validated the themes, drawing on their involvement in the workshop discussions to assess whether the analysis accurately captured how HCI researchers reason about representation, interaction modalities, and perceived limitations when engaging with more-than-human perspectives.


\section{Results}

\subsection{Scenario-Based Discussion}
\label{sec:scenario}

This activity was designed to surface participants’ own representational reasoning across different more-than-human cases. Two groups selected the gum tree (eucalyptus), while one focused on the platypus. The outcomes of these discussions are summarised as comparative scenario vignettes in \autoref{tab:scenario-vignettes}.

Across scenarios, participants differed in how non-human perspectives were approached. Tree-based discussions emphasised indirect and collective representation, drawing on ecological processes, material properties, and long-term environmental change. In contrast, the platypus scenario centred on an individualised perspective, using narrative and experiential framing to reason about lived conditions and behaviours.

\begin{table*}[ht]
\centering
\small
\caption{Comparative scenario vignettes from group discussions.}
\label{tab:scenario-vignettes}
\begin{tabular}{L{1.5cm} L{0.23\linewidth} L{0.29\linewidth} L{0.26\linewidth}}
\toprule
 & \textbf{Gum Tree (Group 1)} & \textbf{Gum Tree (Group 2)} & \textbf{Platypus (Group 3)} \\
\midrule

 &
\begin{minipage}[b]{\linewidth}
    \centering
    \raisebox{-.5\height}{\includegraphics[width=0.5\linewidth]{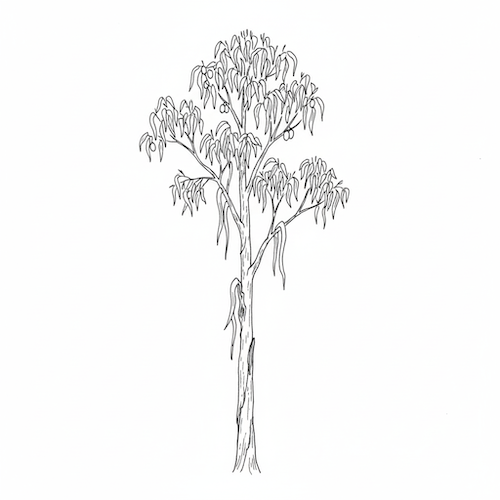}}
\end{minipage}
&
\begin{minipage}[b]{\linewidth}
    \centering
    \raisebox{-.5\height}{\includegraphics[width=0.38\linewidth]{Figures/Gumtree_sketch.png}}
\end{minipage}
&
\begin{minipage}[b]{\linewidth}
    \centering
    \raisebox{-.5\height}{\includegraphics[width=0.6\linewidth]{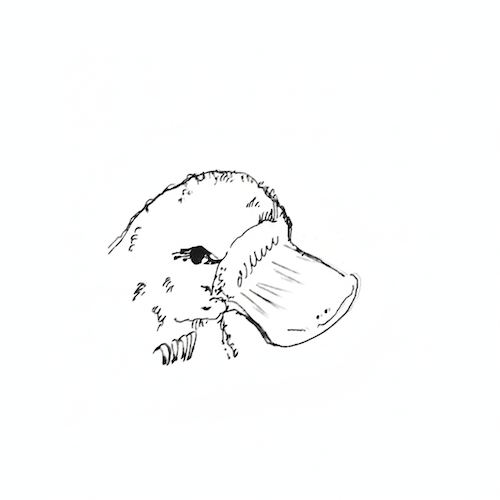}}
\end{minipage}
\\

\textbf{Approach} &
Use mediating representations to enrich understanding of the gum tree’s role, conditions, and impacts over time. &
Articulate indirectly via physical properties (e.g., fuel load, heat, fire behaviour); uses simulation, physical modelling, and absence scenarios to make impacts legible. &
Use an empathetic, narrative, and individualised approach, the platypus is treated as a lived subject.\\

\addlinespace[6pt]

\textbf{Methods referenced} &
Interviews with experts; indigenous knowledge; short stories; timeline-based videos depicting the life of a gum tree; AI simulation; systems mapping; geo-mapping. &
Broadcast media (TV, billboards); hybrid human–non-human representation; fire simulation; immersive `absence' scenarios; systems thinking on resilience and fuel load; speculative scenario modelling. &
Personas; expert representation; animal-mounted cameras; livestreaming; storyboards; journey maps; `day-in-the-life' narratives; route tracking; system mapping; First Nations knowledge.
\\

\bottomrule
\end{tabular}
\end{table*}

\subsection{Voice-Based AI Agent Discussion}
\label{sec:voice}


\paragraph{Voice Qualities and Representational Legitimacy}
Participants treated voice as central to whether the agent could plausibly represent a non-human perspective. Several expressed discomfort with the agent’s vocal delivery, noting that the female-coded voice, conversational style, and American accent made it difficult to perceive the agent as a koala rather than a generic chatbot. Vocal delivery also shaped engagement: the agent’s tone was frequently described as monotonous or lecture-like, leading to reduced attentiveness over time. Participants also noted that repeated responses reinforced the perception of a static, informational voice, and not a dynamic or situational one.

\paragraph{Conversational Flow and the Centralisation of Interaction} 
Participants reflected on how voice-based interaction shaped group dynamics. While speaking to the agent was described as accessible compared to text-based input, its always-listening behaviour introduced interactional friction in group settings. Participants became hesitant to speak while the agent was responding, resulting in pauses and reduced peer-to-peer exchange. As a result, the agent was perceived as central to the interaction, limiting its role as one participant among many.

\paragraph{Education or Advocacy} 
Participants noted that the simple language and explanatory tone made it suitable for educational purposes, but questioned whether this framing was sufficient for advocacy in crisis resilience contexts. Several noted that the agent \textit{`didn’t feel like an advocate,'} characterising its stance as \textit{`more educational than advocacy-oriented.'} Advocacy was characterised as requiring stronger positioning, the capacity to challenge human assumptions, and responsiveness to context. Participants suggested that an advocating non-human voice might need to ask questions, adapt its tone based on the audience and situation, and employ different strategies across settings. 

\paragraph{Plurality and the Limits of a Singular Voice}
Participants questioned whether a single koala could meaningfully represent non-human perspectives in crisis resilience planning, noting that that crises such as bushfires affect populations, families, and multiple species simultaneously. This promoted suggestions that representation might involve multiple koalas with differing experiences, or the inclusion of multiple species within a single interaction. 

\subsection{Immersive Embodied Agent Discussion}
\label{sec:embodied}

\paragraph{Embodiment and Explanatory Value}
Participants emphasised the explanatory value of visual and spatial context when engaging with the embodied koala. Seeing the koala situated within an environment, such as crossing a road,  was described as making scenarios more concrete and easier to reason about than voice-only interaction. Visualisation supported understanding of scale, movement, and spatial relationships, with one participant noting that \textit{`once you can actually see something happening in that space, it makes more sense.'} 

\paragraph{Expectation of Realism}
Participants reflected on how embodiment shaped expectations of realism and behaviour. The koala’s largely static presentation was described as feeling \textit{`like a statue'}, prompting discussion about whether embodied non-human agents should move, respond to gaze, or exhibit species-specific behaviours. 

\paragraph{Authority and the Limits of Personification}
Participants also discussed how embodiment influenced perceptions of authority and legitimacy. One expressed concern about over-personification, cautioning against representations that risk turning animals into \textit{`Disney characters'} or visual gimmicks. Discussion further highlighted that appropriate forms of embodiment were context-dependent. Participants questioned whether a koala should directly speak in all settings, suggesting that mediated forms of representation, such as rangers, narrators, or other authoritative intermediaries, might be more appropriate in certain participatory or educational contexts. 

\subsection{Cross-Cutting Considerations}
\label{sec:crosscutting}

\paragraph{Trust} 
A recurring concern centred on trust, transparency, and factual grounding. Participants questioned how the information provided by an AI-speaking non-human should be validated, asking where the agent’s knowledge comes from and how its claims could be checked. As one participant noted, \textit{`if you’re using AI, it’s going to be very factual-based, how do you know that it’s correct?'} 

\paragraph{Authenticity} 
Participants reflected on how representational choices shape perceptions of authenticity. Beyond immersive or animated representations, participants expressed interest in combining multiple modalities, such as integrating real-world footage with AI- or XR-based representations. Real-world footage was seen as a grounding interpretation and avoiding overly polished or idealised portrayals, particularly in planning contexts where visualisations are often perceived as overly optimistic. Participants also raised concerns about realism and discomfort, questioning whether highly realistic animal representations might produce an \textit{`uncanny'} effect, especially when paired with voice or expressive behaviour.

\paragraph{Motivation over Obligation}
Participants questioned the regulatory framing that dominates participatory planning, where non-human protection is treated as a matter of compliance. Alternative perspectives grounded in cultural obligation and stewardship were discussed as ways of reframing non-human protection as an intrinsic responsibility. In this context, visual, auditory, and immersive elements were seen as capable of supporting affective engagement, complementing factual communication and encouraging reflection and care.

\section{Discussion}

The scenario-based discussions (Section~\ref{sec:scenario}) demonstrated that articulation of non-human perspectives is possible even without voice or embodiment, through indirect and process-oriented forms of representation. Existing literature (e.g., \cite{demir2025rethinking}) has argued that common representation tools in human-centred design, such as personas or journey maps, often fall short in more-than-human contexts due to their grounding in human cognition and bounded agency. Our scenario-based findings both reflect and nuance this critique. Participants rarely applied human-centred representation tools indiscriminately. Instead, they adapted representational strategies to the ecological characteristics and agency of the selected actor. 

The design probe discussions (Sections~\ref{sec:voice}-~\ref{sec:embodied}) revealed that once voice or embodiment is introduced, representation becomes highly sensitive to design choices. Decisions concerning vocal qualities, embodiment, realism, interaction structure, and plurality shaped how authority, advocacy, and legitimacy were perceived. The cross-cutting discussions (Section~\ref{sec:crosscutting}) further situated these concerns within broader issues of trust, factual grounding, motivation, and obligation. Across activities, AI and immersive technologies were framed not merely as communication channels, but as mediators that influence how non-human interests are interpreted and acted upon in participatory planning contexts.

While prior work has identified key parameters for assessing user representation tools in more-than-human design~\citep{demir2025rethinking}, including transferability, depth of representation, multi-perspective integration, and resource investment, our findings suggest that these parameters are necessary but not sufficient. Rather than evaluating representational capacity alone, participants focused on how representations are legitimised and interpreted once they enter decision-making settings. 

Unlike work that positions AI agents themselves as more-than-human entities~\citep{nicenboim2020ai}, our study treats AI systems as representational mediators for non-human species and ecological systems. Despite this difference, both bodies of work share a concern with non-neutral mediation: AI is understood not as a passive interface, but as an active participant in shaping meaning and consequence through interaction.

\section{Limitations and Conclusion}

This study has several limitations. First, the workshop involved HCI researchers, not practitioners directly engaged in crisis resilience planning. While this was a deliberate choice aligned with the paper’s focus, the findings do not capture how representational decisions are negotiated within real-world planning processes or institutional settings. A further study with crisis resilience practitioners is currently in progress to examine how these representational approaches operate within practice contexts. Second, the workshop focused on a limited set of non-human actors and crisis scenarios, which may not reflect the full diversity of ecological contexts or representational challenges.

As a qualitative workshop study, the findings are intended to surface conceptual tensions that can inform future design and empirical investigation. In particular, the study highlights that giving \textit{`voice'} to non-human entities is not a neutral act of translation, but a design challenge involving tensions between legitimacy, authority, and authenticity. Rather than treating more-than-human representation as a problem with a definitive technological solution, this work positions it as an ongoing area of inquiry for HCI research.


\begin{acks}
We thank Ho Yin Chung for technical assistance and insightful discussions that informed the development of the AI voice-based conversational agent, and Hayden Pidgeon for designing and developing the embodied XR design probe used in this workshop. This work was supported by a Collaborative Research Grant from the Sydney Environment Institute and the School of Architecture, Design and Planning, awarded to the project \textit{More-Than-Human Futures: Designing Crisis Resilience Through AI and Immersive Technologies}. We thank the anonymous reviewers for their valuable comments, which have been incorporated into the final version.

The first two authors contributed equally to this work. TTMT and AW co-designed the workshop. AW developed the AI voice-based conversational agent and led workshop facilitation. TTMT led the writing of the first draft and overall manuscript development. AW contributed substantially to the revision of the manuscript. All authors contributed to interpretation and revision of the manuscript.
\end{acks}

\bibliographystyle{ACM-Reference-Format}
\bibliography{references}

@article{coulton2019more,
  title={More-than human centred design: Considering other things},
  author={Coulton, Paul and Lindley, Joseph Galen},
  journal={The Design Journal},
  volume={22},
  number={4},
  pages={463--481},
  year={2019},
  publisher={Taylor \& Francis}
}

@inproceedings{wong2025voice,
author = {Wong, Adrian and Yu, Xinyan and Tran, Tram Thi Minh and Fredericks, Joel},
title = {Voice, Embodiment, and the Relational Roles of AI Agents in Sci-Fi Films: Implications for Representing the More-than-Human},
year = {2025},
isbn = {9798400720161},
publisher = {Association for Computing Machinery},
address = {New York, NY, USA},
url = {https://doi.org/10.1145/3764687.3764691},
doi = {10.1145/3764687.3764691},
booktitle = {Proceedings of the 37th Australian Conference on Human-Computer Interaction},
pages = {36–49},
numpages = {14},
location = {
},
series = {OzCHI '25}
}

@inproceedings{adrian2025ozchiws,
author = {Wong, Adrian and Tran, Tram Thi Minh and Parker, Callum and Hoggenmueller, Marius and Zettna, Nate and Tirado Cortes, Carlos Alfredo and Yoo, Soojeong and Fredericks, Joel},
title = {Engaging Human and Non-Human Perspectives in Crisis Resilience: Designing AI-Supported Immersive Technologies for Inclusive Decision-Making},
year = {2025},
isbn = {9798400720161},
publisher = {Association for Computing Machinery},
address = {New York, NY, USA},
url = {https://doi.org/10.1145/3764687.3767277},
doi = {10.1145/3764687.3767277},
booktitle = {Proceedings of the 37th Australian Conference on Human-Computer Interaction},
pages = {1011–1015},
numpages = {5},
location = {
},
series = {OzCHI '25}
}

@article{frauenberger2019entanglement,
author = {Frauenberger, Christopher},
title = {Entanglement HCI The Next Wave?},
year = {2019},
issue_date = {February 2020},
publisher = {Association for Computing Machinery},
address = {New York, NY, USA},
volume = {27},
number = {1},
issn = {1073-0516},
url = {https://doi.org/10.1145/3364998},
doi = {10.1145/3364998},
journal = {ACM Trans. Comput.-Hum. Interact.},
month = nov,
articleno = {2},
numpages = {27}
}

@inproceedings{usanga2024giving,
author = {Usanga, Chidi and Storni, Cristiano and Light, Ann and Huybrechts, Liesbeth and Teli, Maurizio},
title = {Giving Voice to Nature: Participatory Design with Non-Human Stakeholders for Sustainable Development},
year = {2024},
isbn = {9798400706547},
publisher = {Association for Computing Machinery},
address = {New York, NY, USA},
url = {https://doi.org/10.1145/3661455.3669896},
doi = {10.1145/3661455.3669896},
booktitle = {Proceedings of the Participatory Design Conference 2024: Exploratory Papers and Workshops - Volume 2},
pages = {215–218},
numpages = {4},
location = {Sibu, Malaysia},
series = {PDC '24}
}

@inproceedings{light2024more,
author = {Light, Ann},
title = {More-than-Human Participatory Approaches for Design: Method and Function in Making Relations},
year = {2024},
isbn = {9798400706547},
publisher = {Association for Computing Machinery},
address = {New York, NY, USA},
url = {https://doi.org/10.1145/3661455.3669862},
doi = {10.1145/3661455.3669862},
booktitle = {Proceedings of the Participatory Design Conference 2024: Exploratory Papers and Workshops - Volume 2},
pages = {1–6},
numpages = {6},
location = {Sibu, Malaysia},
series = {PDC '24}
}

@article{nicenboim2023designing,
  author  = {Nicenboim, Iohanna and Giaccardi, Elisa and Redstr{\"o}m, Johan},
  title   = {Designing More-Than-Human AI: Experiments on Situated Conversations and Silences},
  journal = {DIID — Disegno Industriale Industrial Design},
  year    = {2023},
  number  = {80},
  pages   = {12},
  doi     = {10.30682/diid8023c},
  url     = {https://www.diid.it/diid/index.php/diid/article/view/diid80-nicenboim-giaccardi-redstrom}
}

@article{poikolainen2022towards,
  author    = {Poikolainen Rosén, Anton and Normark, Maria and Wiberg, Mikael},
  title     = {Towards More-Than-Human-Centred Design: Learning from Gardening},
  journal   = {International Journal of Design},
  volume    = {16},
  number    = {3},
  pages     = {21--36},
  year      = {2022},
  url       = {https://www.ijdesign.org/index.php/IJDesign/article/view/4402},
  note      = {Available online: \url{https://www.ijdesign.org/index.php/IJDesign/article/view/4402}}
}

@article{tomitsch2025ai,
  author    = {Tomitsch, Martin and Fredericks, Joel and Hoggenmüller, Marius and Crosby, Alexandra and Wong, Adrian and Yu, Xinyan and Huang, Weidong},
  title     = {AI-Supported Participatory Workshops: Middle-Out Engagement for Crisis Events},
  journal   = {Urban Planning},
  volume    = {10},
  number    = {1},
  pages     = {17},
  year      = {2025},
  doi       = {10.17645/up.9165},
  url       = {https://doi.org/10.17645/up.9165},
  issn      = {2183-7635},
  note      = {Open Access under CC BY 4.0}
}

@article{giaccardi2020technology,
author = {Elisa Giaccardi and Johan Redstr\"om},
  title     = {Technology and More-Than-Human Design},
  journal   = {Design Issues},
  volume    = {36},
  number    = {4},
  pages     = {33--44},
  year      = {2020},
  month     = {September},
  issn      = {0747-9360},
  doi       = {10.1162/desi_a_00612},
}

@inproceedings{mancini2023responsible,
author = {Mancini, Clara},
title = {Responsible ACI: Expanding the Influence of Animal-Computer Interaction},
year = {2024},
isbn = {9798400716560},
publisher = {Association for Computing Machinery},
address = {New York, NY, USA},
url = {https://doi.org/10.1145/3637882.3637895},
doi = {10.1145/3637882.3637895},
booktitle = {Proceedings of the Tenth International Conference on Animal-Computer Interaction},
articleno = {3},
numpages = {9},
location = {Raleigh, NC, USA},
series = {ACI '23}
}

@article{roudavski2025dingo,
  title        = {From Dingoes to AI: Who Makes Decisions in More-than-Human Worlds?},
  author       = {Roudavski, Stanislav and Brock, Douglas},
  journal      = {Journal for Human-Animal Studies},
  volume       = {11},
  pages        = {56--96},
  year         = {2025},
  month        = {Mar.},
  doi          = {10.23984/fjhas.145720},
  url          = {https://trace.journal.fi/article/view/145720},
}

@inproceedings{nicenboim2020ai,
author = {Nicenboim, Iohanna and Giaccardi, Elisa and S\o{}ndergaard, Marie Louise Juul and Reddy, Anuradha Venugopal and Strengers, Yolande and Pierce, James and Redstr\"{o}m, Johan},
title = {More-Than-Human Design and AI: In Conversation with Agents},
year = {2020},
isbn = {9781450379878},
publisher = {Association for Computing Machinery},
address = {New York, NY, USA},
url = {https://doi.org/10.1145/3393914.3395912},
doi = {10.1145/3393914.3395912},
booktitle = {Companion Publication of the 2020 ACM Designing Interactive Systems Conference},
pages = {397–400},
numpages = {4},
location = {Eindhoven, Netherlands},
series = {DIS' 20 Companion}
}

@inproceedings{demir2025rethinking,
author = {Demir, Berre Su and Co\c{s}kun, Aykut},
title = {Rethinking Representation in Design: Towards Constructing Parameters for Representation Tools in More-than-Human Design},
year = {2025},
isbn = {9798400714856},
publisher = {Association for Computing Machinery},
address = {New York, NY, USA},
url = {https://doi.org/10.1145/3715336.3735680},
doi = {10.1145/3715336.3735680},
booktitle = {Proceedings of the 2025 ACM Designing Interactive Systems Conference},
pages = {178–194},
numpages = {17},
location = {
},
series = {DIS '25}
}

@inproceedings{Wong2025BeyondHuman,
author = {Wong, Adrian and Spyrou, Julia and Fredericks, Joel},
title = {Beyond Human Interaction: A Contextual Review of Conversational Agents to Represent More-Than-Human Perspectives},
year = {2025},
isbn = {9798400715099},
publisher = {Association for Computing Machinery},
address = {New York, NY, USA},
url = {https://doi.org/10.1145/3726986.3727047},
doi = {10.1145/3726986.3727047},
booktitle = {Proceedings of the 36th Australasian Conference on Human-Computer Interaction},
pages = {810–816},
numpages = {7},
location = {
},
series = {OzCHI '24}
}

@incollection{Beaver2022JustinBeaver,
  author    = {Sierra Rativa, Alexandra and Forero, Aura and Aragon, Arbey and Arias, Nelson and Lopez, Andres and Postma, Marie and Zaanen, Menno},
  title     = {Justin Beaver Stories: A Conversational and Empathic Virtual Animal in Mixed Reality Technology},
  booktitle = {XR Academia},
  publisher = {Open Press TiU},
    address   = {Tilburg, The Netherlands},
  year      = {2022},
  doi       = {10.26116/6wmr-t534}
}

@misc{openai_whisper,
  title        = {Whisper: General-Purpose Speech Recognition Model},
  author       = {{OpenAI}},
  year         = {2023},
  howpublished = {\url{https://github.com/openai/whisper}},
  note         = {Accessed 2026-01-22}
}

@inproceedings{smith2025contemporary,
author = {Smith, Rachel Charlotte and Huybrechts, Liesbeth and Simonsen, Jesper and Loi, Daria},
title = {Contemporary Participatory Design: Research Agendas for Societal Crisis},
year = {2025},
isbn = {9798400720031},
publisher = {Association for Computing Machinery},
address = {New York, NY, USA},
url = {https://doi.org/10.1145/3744169.3744183},
doi = {10.1145/3744169.3744183},
booktitle = {Proceedings of the Sixth Decennial Aarhus Conference: Computing X Crisis},
pages = {182–201},
numpages = {20},
location = {
},
series = {AAR '25}
}

@inproceedings{hirsch2025climate,
author = {Hirsch, Linda and Hwang, Daeun and Johns, Mj and Isbister, Katherine},
title = {HCI for Climate Resilience: Developing an Individual and Community Focused Framework through a Grounded Theory Approach},
year = {2025},
isbn = {9798400714856},
publisher = {Association for Computing Machinery},
address = {New York, NY, USA},
url = {https://doi.org/10.1145/3715336.3735695},
doi = {10.1145/3715336.3735695},
booktitle = {Proceedings of the 2025 ACM Designing Interactive Systems Conference},
pages = {1740–1757},
numpages = {18},
location = {
},
series = {DIS '25}
}

@inproceedings{wallace2013making,
author = {Wallace, Jayne and McCarthy, John and Wright, Peter C. and Olivier, Patrick},
title = {Making design probes work},
year = {2013},
isbn = {9781450318990},
publisher = {Association for Computing Machinery},
address = {New York, NY, USA},
url = {https://doi.org/10.1145/2470654.2466473},
doi = {10.1145/2470654.2466473},
booktitle = {Proceedings of the SIGCHI Conference on Human Factors in Computing Systems},
pages = {3441–3450},
numpages = {10},
location = {Paris, France},
series = {CHI '13}
}

@article{gaver1999design,
author = {Gaver, Bill and Dunne, Tony and Pacenti, Elena},
title = {Design: Cultural probes},
year = {1999},
issue_date = {Jan./Feb. 1999},
publisher = {Association for Computing Machinery},
address = {New York, NY, USA},
volume = {6},
number = {1},
issn = {1072-5520},
url = {https://doi.org/10.1145/291224.291235},
doi = {10.1145/291224.291235},
journal = {Interactions},
month = jan,
pages = {21–29},
numpages = {9}
}

\appendix

\section{Participant Profiles}
\label{appendix:experts}

\begin{table*}[h]
\centering
\small
\renewcommand{\arraystretch}{1.2}
\caption{Participant profiles (P1--P15).}
\label{tab:participant-profiles}
\begin{tabular}{p{0.8cm} p{0.18\textwidth} p{0.65\textwidth}}
\toprule
\textbf{ID} & \textbf{Position / Role} & \textbf{Expertise / Research Focus} \\
\midrule
P1 & Professor & Socio-technical change, urban studies, smart cities, urban transitions \\
P2 & Professor & Human--Technology Interaction, user experience, sustainability-focused HCI \\
P3 & Professor & Smart cities, data care, community advocacy, autonomous mobility, more-than-human futures \\
P4 & Head of School & Interaction design, autonomous mobility, responsible innovation, community-focused design \\
P5 & Associate Head of School & Interaction design, AR/MR, urban computing, more-than-human and multispecies design \\
P6 & Associate Lecturer & Emerging technologies, interaction design \\
P7 & Product Researcher & Digital media, screen design, animation, visual effects, interaction design education \\
P8 & UX Designer & Human-centred design, inclusive and empathetic design practice \\
P9 & PhD Student & Collective mood identification, Internet of Things (IoT), data visualisation \\
P10 & PhD Student & Conversational agents, healthcare AI, co-design with patients and clinicians \\
P11 & PhD Student & Human--Robot Interaction \\
P12 & PhD Student & Design innovation, strategic design \\
P13 & PhD Student & Human--Robot Interaction \\
P14 & PhD Student & Mixed reality, nature interaction, immersive environments \\
P15 & Honours Student & More-than-human design research \\
\bottomrule
\end{tabular}
\end{table*}

\end{document}